\documentclass[12pt] {article}
\usepackage{psfrag}
\usepackage{graphicx}
\usepackage{latexsym,amsfonts}
\usepackage{amssymb}
\begin{document}
\setcounter{page}{1}
\def\theequation{\arabic{section}.\arabic{equation}}
\def\theequation{\thesection.\arabic{equation}}
\setcounter{section}{0}

\title{On the ground state of a free massless (pseudo)scalar field in
two dimensions}

\author{M. Faber\thanks{E--mail: faber@kph.tuwien.ac.at, Tel.:
+43--1--58801--14261, Fax: +43--1--5864203} ~~and~
A. N. Ivanov\thanks{E--mail: ivanov@kph.tuwien.ac.at, Tel.:
+43--1--58801--14261, Fax: +43--1--5864203}~\thanks{Permanent Address:
State Polytechnical University, Department of Nuclear Physics, 195251
St. Petersburg, Russian Federation}}

\date{\today}

\maketitle
\vspace{-0.5in}
\begin{center}
{\it Atominstitut der \"Osterreichischen Universit\"aten,
Arbeitsbereich Kernphysik und Nukleare Astrophysik, Technische
Universit\"at Wien, \\ Wiedner Hauptstr. 8-10, A-1040 Wien,
\"Osterreich }
\end{center}

\begin{center}
\begin{abstract}
We investigate the ground state of a free massless (pseudo)scalar
field in 1+1--dimensional space--time. We argue that in the quantum
field theory of a free massless (pseudo)scalar field without infrared
divergences (Eur. Phys. J. C {\bf 24}, 653 (2002)) the ground state
can be represented by a tensor product of wave functions of the
fiducial vacuum and of the collective zero--mode, describing the
motion of the ``center of mass'' of a free massless (pseudo)scalar
field. We show that the bosonized version of the BCS wave function of
the ground state of the massless Thirring model obtained in
(Phys. Lett. B {\bf 563}, 231 (2003)) describes the ground state of
the free massless (pseudo)scalar field.
\end{abstract}
\end{center}

\newpage

\section{Introduction}
\setcounter{equation}{0}

\hspace{0.2in} The problem which we study in this paper is related to
our investigations of the massless Thirring model\cite{FI1},
where we have found a new phase with a wave function of the BCS--type
and massive quasiparticles. The (pseudo)scalar collective excitations
$\vartheta(x)$ of these massive quasiparticles are bound by a Mexican
hat potential and described by the Lagrangian
\begin{eqnarray}\label{label1.1}
{\cal L}(x) =
\frac{1}{2}\,\partial_{\mu}\vartheta(x)\partial^{\mu}\vartheta(x),
\end{eqnarray}
invariant under field translations
\begin{eqnarray}\label{label1.2}
 \vartheta(x) \to \vartheta\,'(x) = \vartheta(x) + \alpha,
\end{eqnarray}
where $\alpha$ is an arbitrary parameter $\alpha \in \mathbb{R}^1$.
The parameter $\alpha$ is related to the chiral phase $\alpha_{\rm A}$
of chiral rotations of the massless Thirring fermion fields $\alpha =
-2\alpha_{\rm A}$ \cite{FI1,FI2}.

The continuous symmetry (\ref{label1.2}) can be described in terms of
the total charge operator $Q(x^0)$ defined by
\cite{FI2,FI3,FI4}
\begin{eqnarray}\label{label1.3}
Q(x^0) = \int^{+\infty}_{-\infty}dx^1\,j_0(x^0,x^1) =
\int^{+\infty}_{-\infty}dx^1\,\frac{\partial \vartheta(x)}{\partial
x^0} = \int^{+\infty}_{-\infty}dx^1\,\Pi(x^0,x^1),
\end{eqnarray}
where $j_0(x)$ is the time--component of the conserved current
$j_{\mu}(x) = \partial_{\mu}\vartheta(x)$ and $\Pi(x) = j_0(x)$ is the
conjugate momentum of the $\vartheta$--field obeying the canonical
commutation relation
\begin{eqnarray}\label{label1.4}
[\Pi(x^0,x^1),\vartheta(x^0,y^1)] = -i\delta(x^1 - y^1).
\end{eqnarray}
From this canonical commutation relation follows
\begin{eqnarray}\label{label1.5}
\vartheta\,'(x) = e^{\textstyle +i\alpha
Q(x^0)}\,\vartheta(x)\,e^{\textstyle - i\alpha Q(x^0)} = \vartheta(x)
+ \alpha.
\end{eqnarray}
Acting with the operator $e^{\textstyle -i\alpha\,Q(0)}$ on the vacuum
wave function $|\Psi_0\rangle$ we get the wave function
\begin{eqnarray}\label{label1.6}
|\alpha\rangle = e^{\textstyle -i\alpha\,Q(0)}|\Psi_0\rangle.
\end{eqnarray}
This wave function is normalized to unity and possesses all properties
of the vacuum state \cite{KY76}.

The average value of the $\vartheta$--field calculated for the wave
functions $|\alpha\rangle$ is equal to
\begin{eqnarray}\label{label1.7}
\langle\alpha|\vartheta(x)|\alpha\rangle = \alpha.
\end{eqnarray}
The same result can be obtained for the vacuum expectation value
of the $\vartheta\,'$--field (\ref{label1.2})
\begin{eqnarray}\label{label1.8}
\langle \Psi_0|\vartheta\,'(x)|\Psi_0\rangle = \langle
\Psi_0|\vartheta(x)|\Psi_0\rangle + \alpha = \alpha.
\end{eqnarray}
This testifies that the parameter $\alpha$ describes the position of
the ``center of mass''. Hence, it is related to the collective
zero--mode of the free massless (pseudo)scalar field $\vartheta(x)$
\cite{FI2}--\cite{FI4}.

The quantum field theory of the free massless (pseudo)scalar field
$\vartheta(x)$ with the Lagrangian (\ref{label1.1}) is well--defined
if the collective zero--mode, describing the motion of the ``center of
mass'' of the field $\vartheta(x)$, is removed from the states which
can be excited by an external source $J(x)$ in the generating
functional of Green functions \cite{FI2}--\cite{FI4}
\begin{eqnarray}\label{label1.9}
Z[J] &=& \Big\langle \Psi_0\Big|{\rm T}\Big(e^{\textstyle i\int
d^2x\,\vartheta(x)J(x)}\Big)\Big|\Psi_0\Big\rangle =\nonumber\\
&=&\int {\cal D}\vartheta\,e^{\textstyle i\int
d^2x\,\Big[\frac{1}{2}\partial_{\mu}\vartheta(x)
\partial^{\mu}\vartheta(x) + \vartheta(x)J(x)\Big]},
\end{eqnarray}
where ${\rm T}$ is the time--ordering operator. According to the
analysis \cite{FI2,FI3,FI4} the collective
zero--mode cannot be excited by any perturbation of the external
source $J(x)$ if the external source obeys the constraint
\cite{FI2,FI3,FI4}
\begin{eqnarray}\label{label1.10}
\int d^2x\,J(x) = \tilde{J}(0) = 0.
\end{eqnarray}
The same constraint one needs for the perturbative renormalization of
the sine--Gordon model \cite{FI5}. It has been shown that the
sine--Gordon model is the bosonized version of the massive Thirring
model with fermion fields quantized in the chirally broken phase
\cite{FI1}.

As has been shown in\cite{FI2,FI3,FI4} the existence
of the chirally broken phase for the massless Thirring model with a
non--vanishing fermion condensate \cite{FI1} and the
spontaneously broken field--shift symmetry (\ref{label1.2}) of the
free massless (pseudo)scalar field $\vartheta(x)$, characterized by a
non--vanishing spontaneous magnetization
\cite{FI2,FI3,FI4}, does not contradict both the
Mermin--Wagner--Hohenberg theorem \cite{MWH} and Coleman's theorem
\cite{SC73}.  The irrelevance of the Mermin--Wagner--Hohenberg theorem
\cite{MWH} to the problem of the existence of the chirally broken
phase in the massless Thirring model is rather straightforward.
Indeed, the Mermin--Wagner--Hohenberg theorem \cite{MWH}, proved for
non--zero temperature, tells nothing about spontaneous breaking of
continuous symmetry in 1+1--dimensional quantum field theories at
temperature zero\cite{FI2,FI3,FI4}. Since the
chirally broken phase of the massless Thirring model \cite{FI1}
has been found at temperature zero, the Mermin--Wagner--Hohenberg
theorem \cite{MWH} does not suppress this phase.

The absence of spontaneous breaking of continuous symmetry and
Goldstone bosons in 1+1--dimensional quantum field theories at
zero--temperature one connects with Coleman's theorem \cite{SC73}.
Following Wightman's axioms \cite{AW64}, demanding the definition of
Wightman's observables on test functions from the Schwartz class
${\cal S}(\mathbb{R}^{\,2})$, Coleman has argued that there are no
Goldstone bosons, massless (pseudo)scalar fields \cite{SC73}. In turn,
the absence of Goldstone bosons \cite{JG61} can be interpreted as the
absence of spontaneous breaking of continuous symmetry \cite{SC75}.
Coleman's assertion is an extension of the well--known statement of
Wightman \cite{AW64} that a non--trivial quantum field theory of a
free massless (pseudo)scalar field does not exist in 1+1--dimensional
space--time in terms of Wightman's observables defined on the test
functions from ${\cal S}(\mathbb{R}^{\,2})$.

Such a strict conclusion concerning the non--existence of a
1+1--dimensional quantum field theory of a free massless
(pseudo)scalar field $\vartheta(x)$ has been drawn from the
logarithmic divergences of the two--point Wightman functions
\cite{AW64}.

The massless (pseudo)scalar field $\vartheta(x)$ has the following
expansion into plane waves \cite{FI1,FI2,FI3,
FI4}
\begin{eqnarray}\label{label1.11}
\vartheta(x) =
\int^{+\infty}_{-\infty}\frac{dk^1}{2\pi}\,\frac{1}{2k^0}\,
\Big(a(k^1)\,e^{\textstyle -i\,k\cdot x} +
a^{\dagger}(k^1)\,e^{\textstyle i\,k\cdot x}\Big),
\end{eqnarray}
where $a(k^1)$ and $a^{\dagger}(k^1)$ are annihilation and creation
operators and obey the standard commutation relation
\cite{FI1,FI2,FI3,FI4}
\begin{eqnarray}\label{label1.12}
[a(k^1), a^{\dagger}(q^1)] = (2\pi)\,2k^0\,\delta(k^1 - q^1).
\end{eqnarray}
For the free massless (pseudo)scalar field (\ref{label1.11}) one can
define the Wightman function
\begin{eqnarray}\label{label1.13}
D^{(+)}(x; \mu) &=& \langle
\Psi_0|\vartheta(x)\vartheta(0)|\Psi_0\rangle =\nonumber\\
&=&\frac{1}{2\pi}\int^{+\infty}_{-\infty}\frac{dk^1}{2k^0}\,
e^{\textstyle -\,i\,k\cdot x} = - \frac{1}{4\pi}\,{\ell n}[-\mu^2x^2 +
i\,0\cdot\varepsilon(x^0)],
\end{eqnarray}
where $\varepsilon(x^0)$ is the sign function, $x^2 = (x^0)^2 -
(x^1)^2$, $k\cdot x = k^0x^0 - k^1x^1$, $k^0 = |k^1|$ is the energy of
free massless (pseudo)scalar quantum with a momentum $k^1$ and $\mu$
is the infrared cut--off reflecting the infrared divergences of the
Wightman functions (\ref{label1.13}). 

According to Wightman's axioms \cite{AW64} a well--defined quantum
field theory of a free massless (pseudo)scalar field $\vartheta(x)$
should be formulated in terms of Wightman's observables
\begin{eqnarray}\label{label1.14}
 \vartheta(h) = \int d^2x\,h(x)\vartheta(x)
\end{eqnarray}
determined on the test functions $h(x)$ from the Schwartz class ${\cal
S}(\mathbb{R}^{\,2})$ \cite{AW64}. In terms of Wightman's observables
(\ref{label1.14}) one can define a quantum state $|h\rangle$
\cite{AW64}
\begin{eqnarray}\label{label1.15}
|h\rangle = \vartheta(h)|\Psi_0\rangle = \int
 d^2x\,h(x)\vartheta(x)|\Psi_0\rangle,
\end{eqnarray}
where $|\Psi_0\rangle$ is the wave function of the ground state.  The
squared norm of this quantum state is equal to \cite{FI2}
\begin{eqnarray}\label{label1.16}
&&\|h\|^2 = \langle h| h\rangle = \int\!\!\!\int
d^2xd^2y\,h^*(x)D^{(+)}(x -y; \mu)h(y) =
\frac{1}{2\pi}\int^{+\infty}_{-\infty}\frac{dk^1}{2k^0}\,
|\tilde{h}(k^0,k^1)|^2 =\nonumber\\ &&=
\frac{1}{2\pi}\int^{+\infty}_0\frac{dk^1}{k^1}\,
|\tilde{h}(k^1,k^1)|^2 = \frac{1}{2\pi}\lim_{\mu \to
0}\int^{+\infty}_{\mu}\frac{dk^1}{k^1}\, |\tilde{h}(k^1,k^1)|^2 =
-\frac{1}{2\pi}\,|\tilde{h}(0,0)|^2\,\lim_{\mu \to 0}{\ell
n}\,\mu\nonumber\\ && -\frac{1}{2\pi}\int^{+\infty}_0dk^1\,{\ell
n}\,k^1\,\frac{d}{dk^1} |\tilde{h}(k^1,k^1)|^2 =
-\frac{1}{2\pi}\,|\tilde{h}(0,0)|^2\,\lim_{\mu \to 0}{\ell
n}\,\mu\nonumber\\ && +\frac{1}{2\pi}\int^{+\infty}_{-\infty}dk^1\,
\frac{d}{dk^1}[\theta(k^1){\ell n}\,k^1]\,|\tilde{h}(k^1,k^1)|^2,
\end{eqnarray}
where we have used Wightman's formula 
$$
\lim_{\delta \to 0+}\int^{\infty}_{\delta}\frac{\varphi(x)}{x}\,dx = -
\lim_{\delta \to 0+}{\ell n}\,\delta\,\varphi(0) - \int^{\infty}_0
{\ell n}x\,\frac{d\varphi(x)}{dx}\,dx = 
$$
$$
= - \lim_{\delta \to 0+}{\ell n}\,\delta\,\varphi(0) +
\int^{\infty}_{-\infty}\frac{d}{dx}[\theta(x){\ell
n}\,x]\,\varphi(x)\,dx
$$
(see Ref.[25] of Carg$\grave{\rm e}$se Lectures \cite{AW64}).

Since the Fourier transform $\tilde{h}(k^0,k^1)$ of the test function
$h(x)$ from the Schwartz class ${\cal S}(\mathbb{R}^{\,2})$ has a
support at $k^0 = k^1 = 0$, i.e $\tilde{h}(0,0) \neq 0$, the momentum
integral is logarithmically divergent in the infrared region at $\mu
\to 0$. The convergence of the momentum integral in the infrared
region can be provided only for the test functions from the Schwartz
class ${\cal S}_0(\mathbb{R}^{\,2}) = \{h(x) \in {\cal
S}(\mathbb{R}^{\,2}); \tilde{h}(0,0) = 0\}$ \cite{AW64}.

Recently \cite{FI3,FI4} we have analysed the physical
meaning of the test functions $h(x)$ in Wightman's observables
(\ref{label1.14}). We have shown that the test functions can be
interpreted as {\it apparatus functions} characterizing the device
used by the observer for detecting quanta of the free massless
(pseudo)scalar field. This interpretation of test functions agrees
with our results obtained in Ref.\cite{FI2}, where we have shown
that a quantum field theory of a free massless (pseudo)scalar field
$\vartheta(x)$ can be constructed without infrared divergences if one
removes from the $\vartheta$--field the collective zero--mode,
describing the motion of the ``center of mass''. We have shown that
the collective zero--mode does not affect the evolution of the other
modes of the free massless (pseudo)scalar field $\vartheta(x)$. The
removal of the collective zero--mode has been carried out within the
path--integral approach in terms of the generating functional of Green
functions defined by (\ref{label1.9}). As has been shown in
\cite{FI2} the generating functional of Green functions
(\ref{label1.9}) with $\tilde{J}(0) \neq 0$ vanishes identically,
$Z[J] = 0$. This agrees well with Wightman's statement \cite{AW64}
about the non--existence of a quantum field theory of a free massless
(pseudo)scalar field defined on test functions $h(x)$ from ${\cal
S}(\mathbb{R}^{\,2})$ with $\tilde{h}(0,0) \neq 0$.

Hence, the removal of the collective zero--mode of the
$\vartheta(x)$-field implies the immeasurability of this state in
terms of Wightman's observables.  The insensitivity of the detectors
to the collective zero--mode can be obtained by the constraint
$\tilde{h}(0,0) = 0$ \cite{FI3,FI4}. Mathematically this means that
the test functions $h(x)$ should belong to the Schwartz class ${\cal
S}_0(\mathbb{R}^{\,2}) = \{h(x) \in {\cal S}(\mathbb{R}^{\,2});
\tilde{h}(0,0) = 0\}$ \cite{FI2,FI3,FI4}. As has been shown in
\cite{FI3,FI4} the quantum field theory of a free massless
(pseudo)scalar field $\vartheta(x)$ defined on the test functions from
${\cal S}_0(\mathbb{R}^{\,2})$ is unstable under spontaneous breaking
of the continuous symmetry (\ref{label1.2}). Quantitatively the
symmetry broken phase is characterized by a non--vanishing spontaneous
magnetization ${\cal M} = 1$ \cite{FI2,FI3,FI4}.  Goldstone bosons are
the quanta of a free massless (pseudo)scalar field
\cite{FI2,FI3,FI4}. Coleman's theorem reformulated for the test
functions from ${\cal S}_0(\mathbb{R}^{\,2}) = \{h(x) \in {\cal
S}(\mathbb{R}^{\,2}); \tilde{h}(0,0) = 0\}$ does not refute this
statement.

The paper is organized as follows. In Section 2 we describe the
collective zero--mode by a rigid rotor. In Section 3 we calculate the
generating functional of Green functions and show that the infrared
divergences of the free massless (pseudo)scalar field are due to the
classical evolution of the collective zero--mode from infinite past to
infinite future. In Section 4 we construct the wave function of the
ground state of the free massless (pseudo)scalar field for the
bosonized version of the massless Thirring model which is of the
BCS--type.  In the Conclusion we discuss the obtained results.

\section{Collective zero--mode}
\setcounter{equation}{0}

\hspace{0.2in} The collective zero--mode of the free massless
(pseudo)scalar field $\vartheta(x) = \vartheta(x^0,x^1)$, describing
the ``center of mass'' motion, is a mode orthogonal to all vibrational
modes. Due to this we can treat the field $\vartheta(x)$ in the form
of the following decomposition
\begin{eqnarray}\label{label2.1}
\vartheta(x) = \vartheta_0(x^0) + \vartheta_{\rm v}(x) ,
\end{eqnarray}
where $\vartheta_{\rm v}(x)$ is the field of all vibrational
modes. This decomposition can be very well justified for the free
massless (pseudo)scalar field in the finite volume $L$ (see
Eq.(\ref{label2.8})). The Lagrangian of the $\vartheta_{\rm
v}(x)$--field is given by
\begin{eqnarray}\label{label2.3}
{\cal L}_{\rm v}(x) = \frac{1}{2}\Big(\frac{\partial \vartheta_{\rm
v}(x)}{\partial x^0}\Big)^2 - \frac{1}{2} \Big(\frac{\partial
\vartheta_{\rm v}(x)}{\partial x^1}\Big)^2.
\end{eqnarray}

Unlike the oscillator modes of a free massless (pseudo)scalar field
the collective zero--mode $\vartheta_0(x^0)$ is defined by the
Lagrangian
\begin{eqnarray}\label{label2.4}
{\cal L}_0(x^0) = \frac{1}{2}\,\dot{\vartheta}^2_0(x^0),
\end{eqnarray}
which does not have the form of the Lagrangian of a vibrational mode
with a pair of squared terms $(\dot{\vartheta}^2_0 -
\vartheta^{\,'2}_0)/2$, where $\vartheta^{\,'}_0$ is a spatial
derivative of the $\vartheta_0$--field. The Lagrangian of the ``center
of mass'' motion (\ref{label2.4}) does not contain a ``potential
energy'', i.e. $\vartheta^{\,'2}_0/2$, responsible for a restoring
force as in the vibrational modes. Hence, the collective zero--mode
cannot be quantized in terms of annihilation and creation operators
$a(0)$ and $a^{\dagger}(0)$.

Our assertion concerning the decomposition of the free massless
(pseudo)scalar field $\vartheta(x)$ into a collective zero--mode
$\vartheta_0(x^0)$ and vibrational modes $\vartheta_{\rm v}(x)$ can be justified
as follows. As has been shown in \cite{JW80} a free massless
(pseudo)scalar field $\vartheta(x)$ described by the Lagrangian
(\ref{label1.1}) is equivalent to a one--dimensional linear chain of
$N$ oscillators with equal masses, equal equilibrium separations and a
potential energy taking into account only nearest neighbors. Their
motion can be described in terms of displacements $q_i(x^0)\,(i =
1,\ldots,N)$. The Lagrange function can be written as \cite{JW80}
\begin{eqnarray}\label{label2.5}
L(x^0) = \frac{1}{2}\sum^N_{i = 1}\dot{q}^2_i(x^0) +
\frac{1}{2}\sum^{N}_{i>j} (q_i(x^0) - q_j(x^0))^2.
\end{eqnarray}
In normal coordinates $Q_n(x^0)\,(n = 0, 1,\ldots, N - 1)$ the
Lagrange function (\ref{label2.5}) reads
\begin{eqnarray}\label{label2.6}
L(x^0) = \frac{1}{2}\,\dot{Q}^2_0(x^0) + \frac{1}{2}\sum^{N-1}_{n =
1}(\dot{Q}^2_n(x^0) - \omega^2_nQ^2_n(x^0)),
\end{eqnarray}
where $Q_0(x^0)$ is the collective zero--mode, describing the motion
of the ``center of mass'' of the system \cite{FI2}, $Q_n(x^0)$
are vibrational normal modes with frequencies $\omega_n$. In the limit
$N \to \infty$ the Lagrange function (\ref{label2.6}) reduces to the
form \cite{JW80}
\begin{eqnarray}\label{label2.7}
L(x^0) &=& \int^{L/2}_{-L/2}dx^1\,\frac{1}{2}\,\dot{\vartheta}^2_0(x^0)
+ \frac{1}{2} \int^{L/2}_{-L/2}dx^1\,\partial_{\mu}\vartheta_{\mathrm
v}(x) \partial^{\mu}\vartheta_{\rm v}(x)=\nonumber\\
&=&\frac{L}{2}\,\dot{\vartheta}^2_0(x^0) + \frac{1}{2}
\int^{L/2}_{-L/2}dx^1\,\partial_{\mu}\vartheta_{\mathrm
v}(x)\partial^{\mu}\vartheta_{\rm v}(x).
\end{eqnarray}
This is exactly the continuum limit of a one--dimensional chain of $N$
oscillators with nearest neighbour coupling \cite{JW80}.

For finite volume $L$ the discretized form of the $\vartheta$--field
with the expansion of the $\vartheta_{\rm v}$--field into plane
waves reads
\begin{eqnarray}\label{label2.8}
\vartheta(x) = \vartheta_0(x^0) + \sum_{n \in \mathbb{Z}, n \neq
0}\Big(a_n\,e^{\textstyle -ik^0_nx^0 + ik^1_nx^1} +
a^{\dagger}_n\,e^{\textstyle +ik^0_nx^0 - ik^1_nx^1}\,\Big).
\end{eqnarray}
The creation and annihilation operators $a^{\dagger}_n$ and $a_n$ obey
the commutation relations
\begin{eqnarray}\label{label2.9}
{[a_n,a^{\dagger}_{n^{\,'}}]} &=& \delta_{n n^{\,'}},\nonumber\\
{[a^{\dagger}_n,a^{\dagger}_{n^{\,'}}]} &=& {[a_n,a_{n^{\,'}}]} = 0,
\end{eqnarray}
where $k_n = (k^0_n,k^1_n)$ is the 2--dimensional momentum defined by
$k_n = (2\pi|n|/L, 2\pi n /L)$ for $n \in \mathbb{Z}$ and $0 \le
x^1 \le L$. The annihilation operators act on the vacuum state as
$a_n|\Psi_0\rangle = 0$.

The Lagrange function of the collective zero--mode $\vartheta_0(x^0)$
is equal to
\begin{eqnarray}\label{label2.10}
L_0(\vartheta_0,\dot{\vartheta}_0) = \int^{L/2}_{-L/2}dx^1\,{\cal
L}_0(x) =\frac{L}{2}\,\dot{\vartheta}^2_0(x^0).
\end{eqnarray}
Such a Lagrange function(\ref{label2.10}) can be used to describe a
mechanical system, a rigid rotor, for which $\vartheta_0(x^0)$ is a
periodic angle $\vartheta_0(x^0) = \vartheta_0(x^0) + 2\pi\,({\rm
  mod}\,2\pi)$, $\dot{\vartheta}_0(x^0)$ is the angular velocity and
$L$ can be interpreted as the moment of inertia.

The classical equation of motion $\ddot{\vartheta}_0(x^0) = 0$ has the
general solution
\begin{eqnarray}\label{label2.11}
\vartheta_0(x^0) = \Omega_0 x^0 + \alpha,
\end{eqnarray}
where $\Omega_0$ is the angular velocity or the frequency of the
rotation of the rigid rotor.  Setting $\Omega_0 = 0$ we get
$\vartheta_0(x^0) = \alpha$.

The classical conjugate momentum of the collective zero--mode
$\vartheta_0(x^0)$ is equal to
\begin{eqnarray}\label{label2.12}
\pi_0(x^0) = \frac{\partial L_0(\vartheta_0,
\dot{\vartheta}_0)}{\partial \dot{\vartheta}_0} = L\,\dot{\vartheta}_0
\end{eqnarray}
and the Hamilton function is defined by
\begin{eqnarray}\label{label2.13}
h_0(\vartheta_0,\pi_0) = \frac{\pi^2_0(x^0)}{2 L}.
\end{eqnarray}
Substituting (\ref{label2.8}) in (\ref{label1.3}) one can show that
the conjugate momentum $\pi_0(x^0)$ coincides with the total charge
operator $Q(x^0)$, i.e. $Q(x^0) = \pi_0(x^0)$. This agrees with the
analysis of the ground state of the massive Schwinger model by Kogut
and Susskind \cite{KS75}. 

For the quantum mechanical description of the rigid rotor we use the
$\vartheta_0$--representation. In this case the conjugate momentum is
defined by $\hat{\pi}_0 = -i d/d\vartheta_0$ and the Hamilton operator
reads
\begin{eqnarray}\label{label2.14}
\hat{h}_0(\vartheta_0,\hat{\pi}_0) = \frac{\hat{\pi}^2_0(x^0)}{2 L} =
-\frac{1}{2 L}\frac{d^2}{d \vartheta^2_0}.
\end{eqnarray}
The wave function $\psi(\vartheta_0)$ of the collective zero--mode in
the $\vartheta_0$--representation is the solution of the Schr\"odinger
equation
\begin{eqnarray}\label{label2.15}
-\frac{1}{2 L}\frac{d^2\psi(\vartheta_0)}{d \vartheta^2_0} =
E_0\psi(\vartheta_0).
\end{eqnarray}
Imposing periodic boundary conditions $\psi(\vartheta_0) =
\psi(\vartheta_0 + 2\pi)$ the normalized solutions of this equation
read
\begin{eqnarray}\label{label2.16}
\psi_m(\vartheta_0) = \langle \vartheta_0 \mid m \rangle =
\frac{1}{\sqrt{2\pi}}\,e^{\textstyle +i\,m\,\vartheta_0},\quad
m=0,\pm1,\pm 2,\ldots,
\end{eqnarray}
where $m$ is the ``magnetic'' quantum number, $m \in \mathbb{Z}$. The
wave functions $\psi_m(\vartheta_0)$ are also eigenfunctions of the
conjugate momentum $\hat{\pi}_0$ and the total charge operator
$\hat{Q} = \hat{\pi}_0 = -i d/d\vartheta_0$ with the eigenvalues $m
\in \mathbb{Z}$
\begin{eqnarray}\label{label2.17}
Q\psi_m(\vartheta_0) = m\,\psi_m(\vartheta_0) \Longleftrightarrow
\hat{Q}|m\rangle = m|m\rangle.
\end{eqnarray}
The energy spectrum is defined by
\begin{eqnarray}\label{label2.18}
E^{(m)}_0 = \frac{m^2}{2L}.
\end{eqnarray}

The wave function of the free massless (pseudo)scalar field
$\vartheta(x)$ can be represented in the form of direct product of the
collective zero--mode and vibrational modes
\begin{eqnarray}\label{label2.19}
|\Psi\rangle = |m\rangle \otimes |n_1\rangle_{_1} \otimes |n_2\rangle_{_2} \otimes  \cdots \otimes |n_k \rangle_{_k} \otimes \cdots,
\end{eqnarray}
where $|m\rangle$ is the state of the zero--mode and $|n_k
\rangle_{_k}$ is the wave function for the $k$-th
vibrational mode with $n_k$ quanta.

The total Hamilton and momentum operators of the free
massless (pseudo)scalar field $\vartheta(x)$, defined by
(\ref{label2.8}), is equal to
\begin{eqnarray}\label{label2.20}
\hat{h}[\vartheta] &=& \frac{\hat{\pi}^2_0}{2L} +
\frac{2\pi}{L}\sum_{n \in \mathbb{Z}}|n|\,a^{\dagger}_n
a_n,\nonumber\\ \hat{\pi}[\vartheta] &=& \hat{\pi}_0 +
\frac{2\pi}{L}\sum_{n \in \mathbb{Z}}n\,a^{\dagger}_n a_n.
\end{eqnarray}
It is well--known that the wave function of the ground state should be
eigenfunction of the total Hamilton and momentum operators with
eigenvalue zero. For finite $L$ this requirement is fulfilled only for
the wave function
\begin{eqnarray}\label{label2.21}
|\Omega_0\rangle = |0\rangle_0 \otimes |\Psi_0\rangle,\qquad |\Psi_0\rangle=|0\rangle_{_1} \cdots \otimes |0 \rangle_{_k} \otimes \cdots
\end{eqnarray}
where $|0\rangle$ is the eigenfunction of the operator
(\ref{label2.14}) with eigenvalue $m = 0$. In the
$\vartheta_0$--representation the wave function $|0\rangle$ is equal
to $\vspace{1mm}\vspace{-1mm}\langle \vartheta_0 \mid 0
\rangle = \psi_0(\vartheta_0) = 1/\sqrt{2\pi}$.

Now we can show that the infrared divergences of the free massless
(pseudo)scalar field theory are the quantum field theoretical problem
and they are not the problem at all. Indeed, in reality these quantum
field theoretic divergences are related to the classical motion of the
collective zero--mode from the infinite past at $x^0 = - \infty$ to the
infinite future at $x^0 = + \infty$.

\section{Generating functional of Green functions and the nature of 
infrared divergences } 
\setcounter{equation}{0}

\hspace{0.2in} The generating functional of Green functions for the
free massless (pseudo)scalar field is defined by (\ref{label2.20}) and
reads
\begin{eqnarray}\label{label3.1}
{\cal Z}[J] &=& \langle\Omega_0|{\rm T}\Big(e^{\textstyle i\int
d^2x\,\vartheta(x)J(x)}\Big)|\Omega_0\rangle = \lim_{L,T \to \infty}
Z_0[J_0;L,T] \nonumber\\&&\times\,\langle\Psi_0|{\rm
T}\Big(e^{\textstyle i\int d^2x\, \vartheta_{\rm
v}(x)J(x)}\Big)|\Psi_0\rangle = \lim_{L,T \to \infty}Z_0[J_0;L,T] \;
Z[J].
\end{eqnarray}
The factor $Z[J]$ in this product concerns the vibrational modes and
coincides with (\ref{label1.9}). The other factor
\begin{eqnarray}\label{label3.2}
Z_0[J_0;L,T] &=& e^{\textstyle iW_0[J_0;L,T]}=\nonumber\\
&=&N^{-1}(L,T)\int {\cal D}\vartheta_0\,
\exp\Big\{i\int^T_{-T}dx^0\Big[\frac{L}{2}\, \dot{\vartheta}^2_0(x^0)
+ \vartheta_0(x^0)J_0(x^0)\Big]\Big\}
\end{eqnarray}
is the generating functional for the collective zero--mode, which we
describe as the motion of a rigid rotor; the external source
$J_0(x^0)$ is the integral of the external source $J(x) = J(x^0, x^1)$
over $x^1 \in \mathbb{R}^{\,1}$. The normalization factor $N(L,T)$ is
the inverse of the path integral without external sources
\begin{eqnarray}\label{label3.3}
N(L,T)=\int {\cal D}\vartheta_0\, \exp\Big\{i\frac{L}{2}\int^{+
T}_{-T}dx^0\, \dot{\vartheta}^2_0(x^0) \Big\}
\end{eqnarray}
Let us perform a Fourier transformation
\begin{eqnarray}\label{label3.4}
\vartheta_0(x^0) = \int^{+\infty}_{-\infty}\frac{d\omega}{2\pi}\,
\tilde{\vartheta}_0(\omega)\,e^{\textstyle\,-i\omega x^0}
\end{eqnarray}
and get
\begin{eqnarray}\label{label3.5}
Z_0[J_0; L, T] &=& N^{-1}(L) \int {\cal D}\tilde{\vartheta}_0\,
\exp\Big\{iL\int^{+\infty}_{-\infty}
\frac{d\omega}{2\pi}\int^{+\infty}_{-\infty}
\frac{d\omega\,'}{2\pi}\,\frac{\sin((\omega + \omega\,'\,)T)}{(\omega
+ \omega\,'\,)}\nonumber\\ &&\times\,\Big[-\,\omega\,
\omega\,'\,\tilde{\vartheta}_0
(\omega)\tilde{\vartheta}_0(\omega\,'\,)
+\frac{2}{L}\,\tilde{\vartheta}_0(\omega)
\tilde{J}_0(\omega\,'\,)\Big]\Big\}.
\end{eqnarray}
By quadratic extension $\tilde{\vartheta}_0(\omega) =
\tilde{\varphi}(\omega) - \tilde{J}(\omega)/\omega^2L$ we reduce this
to the form 
\begin{eqnarray}\label{label3.6}
Z_0[J_0; L, T] &=& e^{\textstyle\,iW_0[J_0; L, T]} =\nonumber\\
&=&\exp\Big\{\frac{i}{L} \int^{+\infty}_{-\infty}\frac{d\omega}{2\pi}
\int^{+\infty}_{-\infty}\frac{d\omega\,'}{2\pi}\frac{\sin((\omega +
\omega\,'\,)T)}{(\omega + \omega\,'\,)}\frac{\tilde{J}_0(\omega)
\tilde{J}_0(\omega\,'\,)}{\omega\, \omega\,'\,}\Big\}.
\end{eqnarray}
At finite $T$ the functional $W_0[J_0; L, T]$ has a superficial
infrared divergence (\ref{label1.1}). It seems that due to the
infrared divergence ${\cal W}[J_0; L, T]$ becomes infinite and
$Z_0[J_0; L, T]$ vanishes. This should agree with Wightman's assertion
concerning the non--existence of a well--defined quantum field theory
of a free massless (pseudo)scalar field in 1+1--dimensional
space--time which we discussed above.  In our treatment of the
collective zero--mode of the free massless (pseudo)scalar field the
infrared divergence appears at the quantum mechanical level and admits
a simple physical interpretation, as we will discuss below. The
divergence does not appear due to unbounded quantum fluctuations. It
is just the opposite, there is only one contribution to the generating
functional $Z_0[J_0; L, T]$, the classical trajectory which is defined
by the initial conditions. The correlation functions gets unbounded
due to the laws of classical mechanics.

In order to show that the functional $W_0[J_0; L, T]$ tends to
infinity at $T \to \infty$ and that this is related to a classical
motion we suggest to determine the path--integral (\ref{label3.2}),
decomposing $\vartheta_0(x^0)$ into a classical part
$\bar{\vartheta}_0(x^0)$ and a fluctuating part $\varphi(x^0)$,
$\vartheta_0(x^0) = \bar{\vartheta}_0(x^0) + \varphi(x^0)$. With the
standard conditions $\varphi(-T) = \varphi(+T) = 0$ and choosing the
classical field $\bar{\vartheta}_0(x^0)$ as a special solution of the
``2nd axiom of Newton''
\begin{eqnarray}\label{label3.7}
\ddot{\bar{\vartheta}}_0(x^0) = \frac{1}{L}\,J_0(x^0)
\end{eqnarray}
we get a decoupling of the quantum fluctuations $\varphi(x^0)$.  The
rigid rotor behaves completely classical, the generating functional
depends only on the classical field $\bar{\vartheta}_0(x^0)$ and the
action has the same shape as in (\ref{label3.2}) for the original
field $\vartheta_0$
\begin{eqnarray}\label{label3.8}
Z_0[J_0;L,T] = e^{\textstyle iW_0[J_0;L,T]}
=\exp\Big\{i\int_{-T}^Tdx^0\Big[\frac{L}{2}\,
\dot{\bar{\vartheta}}^2_0(x^0) +
\bar{\vartheta}_0(x^0)J_0(x^0))\Big]\Big\}.
\end{eqnarray}
It is important to emphasize that due to this decoupling of quantum
and classical degrees of freedom the external source $J_0(x^0)$ can
excite the classical degrees of freedom $\vartheta_0$ only.

Integrating by parts in the exponent of (\ref{label3.8}) and using
(\ref{label3.7}) we get for the action integral
\begin{eqnarray}\label{label3.9}
W_0[J_0;L,T] = \frac{1}{2}\int_{-T}^{+T}dx^0\,
\bar{\vartheta}_0(x^0)J_0(x^0)+
\frac{L}{2}\Big[\bar{\vartheta}_0(x^0)
\dot{\bar{\vartheta}}_0(x^0)\Big]_{-T}^{+T}.
\end{eqnarray}
Obviously, this integral and therefore the generating functional
(\ref{label3.2}) depends on the initial condition. This is the usual
situation for classical systems. Now we insert in
Eq.~(\ref{label3.9}) a solution of Eq.~(\ref{label3.7}). The general
solution of (\ref{label3.7}) reads
\begin{eqnarray}\label{label3.10}
\bar{\vartheta}_0(x^0) = \frac{1}{2L}
\int_{-T}^{+T}dy^0|x^0 - y^0| J_0(y^0) + C_1x^0 + C_2.
\end{eqnarray}
For the initial conditions $\bar{\vartheta}_0(-T) =
\dot{\bar{\vartheta}}_0(-T) = 0$ the integration constants are
\begin{eqnarray}\label{label3.11}
C_1 &=& \frac{1}{2L}\int_{-T}^{+T}dy^0 J_0(y^0),\nonumber\\
C_2&=&-\frac{1}{2L}\int_{-T}^{+T}dy^0 y^0 J_0(y^0),
\end{eqnarray}
which lead to a special solution of (\ref{label3.9}) 
\begin{eqnarray}\label{label3.12}
\bar{\vartheta}_0(x^0)
= \frac{1}{2L} \int_{-T}^{+T}dy^0\,(|x^0 - y^0| + (x^0 - y^0)) J_0(y^0)
= \frac{1}{L} \int_{-T}^{x^0} dy^0\,(x^0 - y^0) J_0(y^0).
\end{eqnarray}
Substituting this expression into (\ref{label3.9}) we get
\begin{eqnarray}\label{label3.13}
W_0[J_0;L,T] &=&\frac{1}{4L}\int_{-T}^{+T}dx^0\int^{+ T}_{-T}dy^0
J_0(x^0)|x^0 - y^0|J_0(y^0)\nonumber\\ 
&+&\frac{1}{2L} \int_{-T}^{+T}dy^0\,(T -
y^0)J_0(y^0) \,\int_{-T}^{+T}dx^0J_0(x^0).
\end{eqnarray}
For $\tilde{J}(0) \neq 0$ the functional $W_0[J_0;L,T]$ increases
with the time $T$. This gives a strongly oscillating phase of the
generating functional $Z_0[J_0;L,T]$ providing its vanishing in
the limit $T \to \infty$. This occurs even if the external source
generates an arbitrary small fluctuation of the collective
zero--mode.

This can also be seen defining the classical field
$\langle \vartheta_0(x^0) \rangle$ in terms of the generating
functional $Z_0[J_0;L,T]$.  According to the standard definition
the classical field $\langle \vartheta_0(x^0)\rangle$ is given by
\begin{eqnarray}\label{label3.14}
\langle \vartheta_0(x^0)\rangle
= \frac{1}{i}\frac{\delta {\ell n}Z_0[J_0;L,T]}{\delta J_0(x^0)}
= \frac{\delta W_0[J_0;L,T]}{\delta J_0(x^0)}.
\end{eqnarray}
This defines the linear response of the system to an external force
\begin{eqnarray}\label{label3.15}
\langle \vartheta_0(x^0)\rangle 
=\frac{1}{2L}\int_{-T}^{+T}dy^0
[|x^0 - y^0| + (T-x^0) +(T-y^0) ] J_0(y^0),
\end{eqnarray}
but the response is quadratic
\begin{eqnarray}\label{label3.16}
W_0[J_0;L,T]
= \frac{1}{2}\int_{-T}^{+T}dx^0
J_0(x^0) \langle \vartheta_0(x^0)\rangle,
\end{eqnarray}
as it is well-known for free motion.

In the limit $T\to \infty$ the response $\langle \vartheta_0(x^0)
\rangle$ does not vanish for infinitesimally small external
perturbation. From this behaviour we can conclude that the divergence
prohibiting a quantum field theoretic description of a free massless
(pseudo)scalar field [9], corresponds to a quite natural motion which
is well--known in classical and quantum mechanics of free particles
and rigid rotors. An infinitely small kick at $- T=-\infty$ can lead
to finite translations or rotational angles at $T$ as shown by
(\ref{label3.14}).

The correlation
\begin{eqnarray}\label{label3.17}
\langle \vartheta_0(x^0) \vartheta_0(y^0) \rangle
= \frac{\delta^2 W_0[J_0;L,T]}{\delta J_0(x^0)\delta J_0(y^0)}
=\frac{1}{2L}[|x^0 - y^0| + (T-x^0) +(T-y^0)]
\end{eqnarray}
diverges also with $T$ indicating that the system is not stabilized by
any potential. Only under the constraint $\tilde{J}(0,0) = 0$,
agreeing well with the definition of the quantum field theory of the
free massless (pseudo)scalar field on the Schwartz class ${\cal S}_0$,
the correlation (\ref{label3.17}) remains finite at $\tilde{J}_0(0) = 0$
\begin{eqnarray}\label{label3.18}
\langle \vartheta_0(x^0) \vartheta_0(y^0) \rangle
=\frac{1}{2L} |x^0 - y^0| 
\end{eqnarray}
Since at $\tilde{J}_0(0) = 0$
\begin{eqnarray}\label{label3.19}
\langle \vartheta_0(x^0)\rangle =\frac{1}{2L}\int_{-T}^{+T}dy^0 |x^0 -
y^0| J_0(y^0),
\end{eqnarray}
the response $\langle \vartheta_0(x^0)\rangle$ vanishes for
$J_0(y^0)=0$ and Eq.~(\ref{label3.18}) shows that the center of mass
motion does not decorrelate for large time differences $|x^0 - y^0|$.

The above discussion demonstrates that the divergence of the
correlation (\ref{label3.17}) is not due to large quantum
fluctuations, it is due to the the sensitivity of a free system for
external perturbations. The center of mass motion does not show any
fluctuations, it evolves along the classical trajectory only, see
Eq.~(\ref{label3.8}). This explains also why the correlation does not
vanish for large time intervals.

In the Schwinger formulation of the quantum field theory \cite{SW70}
the generating functional $Z_0[J_0; L, T=\infty]$ defines the
amplitude for the transition from the ground state of the center of
mass motion at $x^0 = -\infty$ to the ground state at $x^0 = +\infty$
caused by the external force $J_0(x^0)$. For vanishing perturbation
this amplitude does not converge to unity and gives the impression
that the evolution of the system is ill defined. But this is not the
case. Let the state of the center of mass at $x^0=0$ be described by
$\vartheta_0=\alpha$ and $\dot{\vartheta}_0 = 0$. Then the classical
evolution of the system guarantees that the system will remain in this
state for any finite time and the corresponding transition amplitude
is unity.

Due to the quadratic extension of the exponent $\vartheta_0(x^0) =
\bar{\vartheta}_0(x^0) + \varphi(x^0)$ in the generating functional of
Green functions $Z_0[J_0; L,T]$ and the condition (\ref{label3.7}),
the vibrational degrees of freedom do not couple to the external
source $J_0(x^0)$ which excites the collective zero--mode.  As a
result, the generating functional of Green functions $Z_0[J_0; L,T]$
is defined by one classical trajectory (\ref{label3.8}) and for
$J_0(x^0) = 0$ depends on the temporal boundary conditions only
\begin{eqnarray}\label{label3.20}
W_0[J_0;L,T] =
\frac{L}{2}\Big[\bar{\vartheta}_0(x^0)
\dot{\bar{\vartheta}}_0(x^0)\Big]_{-T}^{+T}.
\end{eqnarray}
We can take into account $2\pi$-periodicity of $\vartheta_0$ assuming
that the paths $\bar{\vartheta}_0(+ T) - \bar{\vartheta}_0(-T) =
\Delta$ and $\bar{\vartheta}_0(+ T) - \bar{\vartheta}_0(-T) = \Delta +
2\pi m$, where $m \in \mathbb{Z}$, are equivalent and
indistinguishable. Then, we obtain the generating functional
$Z_0[0;L,T]$ in the following form
\begin{eqnarray}\label{label3.21}
\hspace{-0.5in}Z_0[0;L,T] &=& \sqrt{\frac{L}{4\pi i T}}\sum_{m\in
\mathbb{Z}}\exp\Big(\,i\,\frac{L}{4T}\,[\Delta + 2\pi m]^2\Big) =
\nonumber\\
\hspace{-0.5in}&=& \sqrt{\frac{L}{4\pi i T}}\,
\exp\Big(\,i\,\frac{L\Delta^2}{4T}\Big) \,\theta_3 \Big(\frac{\pi
L\Delta}{2T},\frac{\pi L}{T}\Big),
\end{eqnarray}
where $\theta_3(z,t)$ is the Jacobi theta--function defined by
\cite{SH81}
\begin{eqnarray}\label{label3.22}
\theta_3(z,t) = \sum_{m \in \mathbb{Z}}e^{\textstyle\,i\pi\,t\,m^2 +
i\,2\,m\,z}
\end{eqnarray}
with $z = \pi L\Delta/2T$ and $t = \pi L/T$.  The normalization factor
\cite{SH81} in front of the sum over equivalent paths is chosen in
such a way that the path integral for given initial position
$\vartheta_0(-T)$ and arbitrary final position $\vartheta_0(+T)$ gives
unity.  Using the property of the Jacobi theta--function
\cite{SH81}
\begin{eqnarray}\label{label3.23}
\theta_3(z,t) = \sqrt{\frac{i}{t}}\,\exp\Big(-\frac{iz^2}{\pi
t}\Big)\,\theta_3\Big(\frac{z}{t}, - \frac{1}{t}\Big)
\end{eqnarray}
we transcribe the functional $Z_0[0;L,T]$ into the form
\begin{eqnarray}\label{label3.24}
Z_0[0;L,T] &=& \frac{1}{2\pi}\,\theta_3\Big(\frac{\Delta}{2},-
\frac{T}{\pi L}\Big) = \frac{1}{2\pi}\sum_{m \in
\mathbb{Z}}\exp\Big(i\,m\,\Delta -
i\,\frac{T}{L}\,m^2\Big)=\nonumber\\ &=&\frac{1}{2\pi}\sum_{m \in
\mathbb{Z}}e^{\textstyle \,i\,m\,[\vartheta_0(+T) -
\vartheta_0(-T)] - i\,E_m\,2T}.
\end{eqnarray}
The r.h.s. contains a sum over the stationary quantum states of the
rigid rotor \cite{SH81}
\begin{eqnarray}\label{label3.25}
Z_0[0;L,T] = \sum_{m \in \mathbb{Z}}{\langle \vartheta_0(+T)|m
\rangle}{\langle m|\vartheta_0(-T)\rangle} =
{\langle \vartheta_0(+T)|\vartheta_0(-T)\rangle}
\end{eqnarray}
describing the amplitude for the transition $\vartheta_0(-T) \to
\vartheta_0(+T)$.  For the reduction of the r.h.s. of
(\ref{label3.25}) we have used the notations
\begin{eqnarray}\label{label3.26}
{\langle\vartheta_0(+T)|m\rangle} &=&
\frac{1}{\sqrt{2\pi}}\,e^{\textstyle + i\,m\,\vartheta_0(+T) -
i\,E_m\,T},\nonumber\\ {\langle m|\vartheta_0(-T)\rangle}
&=&\frac{1}{\sqrt{2\pi}}\,e^{\textstyle - i\,m\,\vartheta_0(-T) +
i\,E_m\,(-T)}
\end{eqnarray}
with $E_m = m^2/2L$ and $|\vartheta_0(x^0)\rangle$ as eigenstate of
the Heisenberg operator $\hat{\vartheta}_0(x^0)$ and the completeness
condition
\begin{eqnarray}\label{label3.27}
\sum_{m \in \mathbb{Z}}{|m\rangle} {\langle m|} = 1.
\end{eqnarray}
In the limit $T \to \infty$ the excited stated with $m \neq 0$ in the
sum (\ref{label3.24}) are dying out and only the ground state with
$m=0$ survives. This yields $Z_0[0;L,\infty] = 1$.

\section{Wave function of the ground state of the free massless 
(pseudo)scalar field}
\setcounter{equation}{0}

\hspace{0.2in} As has been shown in \cite{FI6} the wave function
of the ground state of the free massless (pseudo)scalar field,
describing the bosonized version of the massless Thirring model,
quantized in the chirally broken phase with the BCS wave function of
the ground state, takes the form
\begin{eqnarray}\label{label4.1}
|\Omega(0)\rangle_{BCS} = \exp\Big(i\,\frac{\pi}{2}\,\frac{M}{g}
\int^{+\infty}_{-\infty}dx^1\,\sin(\beta\hat{\vartheta}(0,x^1))\Big)
|\Psi_0\rangle.
\end{eqnarray}
where $M$ is the dynamical mass of the massless Thirring fermion field
quantized in the chiral broken phase, and $g$ is the Thirring coupling
constant \cite{FI1}. The parameter $\beta$ in the definition of
the wave function (\ref{label4.1}), used in \cite{FI6},
$\sin\beta\vartheta(0,x^1)$, can be removed rescaling the
$\vartheta$--field. Indeed, the Lagrangian (\ref{label2.9}) can be
transcribed as follows
\begin{eqnarray}\label{label4.2}
L_0(\vartheta_0,\dot{\vartheta}_0) =
\frac{L}{2\beta^2}\,(\beta\dot{\vartheta}_0(x^0))^2.
\end{eqnarray}
The moment of inertia is now defined as $L/\beta^2$. The Hamilton
operator reads
\begin{eqnarray}\label{label4.3}
\hat{h}_0(\vartheta_0,\hat{\pi}_0) = -\frac{\beta^2}{2 L}\frac{d^2}{d
\vartheta^2_0}.
\end{eqnarray}
The energy spectrum $E^{(m)}_0$ is given by
\begin{eqnarray}\label{label4.4}
E^{(m)}_0 = \frac{\beta^2}{2L}\,m^2.
\end{eqnarray}
For the $\vartheta_0$--field with the moment of inertia $L/\beta^2$
the total Hamilton and momentum operators are defined by
\begin{eqnarray}\label{label4.5}
\hat{h}[\vartheta] &=& \frac{\beta^2}{2L}\,\hat{\pi}^2_0 +
\frac{2\pi}{L}\sum_{n \in \mathbb{Z}}|n|\,a^{\dagger}_n a_n,\nonumber\\
\hat{\pi}[\vartheta] &=& \hat{\pi}_0 +
\frac{2\pi}{L}\sum_{n \in \mathbb{Z}}n\,a^{\dagger}_n a_n.
\end{eqnarray}
Using the discretized form for the $\vartheta$--field (\ref{label2.8})
the wave function (\ref{label4.1}) can be transcribed into the form
\begin{eqnarray}\label{label4.6}
|\Omega(0)\rangle_{BCS} = e^{\textstyle +\,i\,\lambda\,\sin
\hat{\vartheta}_0}\,|\Omega_0\rangle = e^{\textstyle
+\,i\,\lambda\,\sin \hat{\vartheta}_0}\,|0\rangle \otimes
|\Psi_0\rangle.
\end{eqnarray}
where we have denoted $\lambda = \pi M L/2g$. The r.h.s. of
(\ref{label4.6}) should be taken in the limit $\lambda \to \infty$
that corresponds to $L \to \infty$. It is obvious that the wave
function (\ref{label4.6}) is not invariant under the symmetry
transformations (\ref{label1.2}). Hence, it should describe the ground
state of the free massless (pseudo)scalar field $\vartheta(x)$ in the
symmetry broken phase. The contribution of the $\vartheta_{\rm v}$--field is of
order of $O(1/L)$ and smaller compared with the contribution of the
zero--mode. It can be dropped in the limit $L \to \infty$.  In the
$\vartheta_0$--representation the wave function (\ref{label4.6})
reads
\begin{eqnarray}\label{label4.7}
{\langle \vartheta_0|\Omega(0)\rangle_{BCS}} =
\frac{1}{\sqrt{2\pi}}\,e^{\textstyle
+\,i\,\lambda\,\sin\vartheta_0}\otimes |\Psi_0\rangle.
\end{eqnarray}
The wave function (\ref{label4.7}) can be expanded into the
eigenfunctions (\ref{label2.16}) of the Hamilton operator
(\ref{label2.15}). The result reads
\begin{eqnarray}\label{label4.8}
|\Omega(0)\rangle_{BCS} = \sum^{\infty}_{m = -
\infty}J_m(\lambda)\,|m\rangle\otimes |\Psi_0\rangle,
\end{eqnarray}
where $J_m(\lambda)$ are Bessel functions \cite{AS72} and the limit
$\lambda \to \infty$ is assumed. The normalization of the wave
function (\ref{label4.8}) to unity is caused by \cite{GR65}
\begin{eqnarray}\label{label4.9}
\sum^{\infty}_{m = - \infty}J^2_m(\lambda) = 1.
\end{eqnarray}
Under field--shifts (\ref{label1.2}) the wave function
(\ref{label4.8}) transforms into the wave function
\begin{eqnarray}\label{label4.10}
|\Omega(\alpha)\rangle_{BCS} = \sum^{\infty}_{m = -
\infty}J_m(\lambda)\,e^{\textstyle \,i m\alpha}\,|m\rangle\otimes
|\Psi_0\rangle
\end{eqnarray}
at $\lambda \to \infty$.  The orthogonality relation for
$\alpha\,'\neq \alpha$ is defined by
\begin{eqnarray}\label{label4.11}
_{BCS}\langle\Omega(\alpha\,'\,)|\Omega(\alpha)\rangle_{BCS} &=&
\lim_{\lambda \to \infty}\sum^{\infty}_{m =
-\infty}J^2_m(\lambda)\,e^{\textstyle - im(\alpha\,' -
\alpha)}=\nonumber\\ &=& \lim_{\lambda \to
\infty}J_0\Big(2\lambda\,\sin\Big(\frac{\alpha\,' -
\alpha}{2}\Big)\Big) = \delta_{\alpha\,'\alpha},
\end{eqnarray}
where we used the formula \cite{AS72a}
\begin{eqnarray}\label{label4.12}
\sum^{\infty}_{m = - \infty}J^2_m(\lambda)\,e^{\textstyle -
im(\alpha\,' - \alpha)} = J_0\Big(2\lambda\,\sin\Big(\frac{\alpha\,' -
\alpha}{2}\Big)\Big).
\end{eqnarray}
Now let us show that the BCS wave function (\ref{label4.7}) describes
a quantum state with energy zero.

For this aim we notice that the wave--function (\ref{label4.7}) is not
an eigenfunction of the Hamilton operator (\ref{label4.3}) and the
momentum operator $\hat{\pi}_0 = - i d/d\vartheta_0$. It is
well--known that the wave--function of the ground state should be the
eigenfunction of the Hamilton operator of the quantum field under
consideration with eigenvalue zero \cite{AW64}. Let us show that this
requirement can be satisfied for the BCS wave function
(\ref{label4.7}) by a canonical transformation \cite{AN94}. First, we
act with the operator $\hat{\pi}_0 = - i d/d\vartheta_0$ on the wave
function (\ref{label4.7}) and get
\begin{eqnarray}\label{label4.13}
\Big(- i \frac{d}{d\vartheta_0} -
\lambda\,\cos\vartheta_0\Big){\langle
\vartheta_0|\Omega(0)\rangle_{BCS}} = 0.
\end{eqnarray}
The operator in the l.h.s.\ of (\ref{label4.13}) is the conjugate
momentum operator $\hat{\Pi}_0$ in the $\vartheta_0$--representation
related to the conjugate momentum $\hat{\pi}_0 = - i d/d\vartheta_0$
by the unitary transformation \cite{AN94}
\begin{eqnarray}\label{label4.14}
\hat{\Pi}_0 =
U\,\hat{\pi}_0\,U^{\dagger},
\end{eqnarray}
where $U$ is defined by
\begin{eqnarray}\label{label4.15}
U = e^{\textstyle
+i\,\lambda\,\sin\hat{\vartheta}_0}.
\end{eqnarray}
The unitary operator $U$ relates the wave functions
$|\Omega_0\rangle$ and $|\Omega(0)\rangle_{BCS}$
\begin{eqnarray}\label{label4.16}
|\Omega(0)\rangle_{BCS} = U\,|\Omega_0\rangle.
\end{eqnarray}
The operator $\hat{\Pi}_0$ is equal to
\begin{eqnarray}\label{label4.17}
\hat{\Pi}_0 = \hat{\pi}_0 +
i\lambda\,[\sin\hat{\vartheta}_0,\hat{\pi}_0] = \hat{\pi}_0 -
\lambda\,\cos\hat{\vartheta}_0,
\end{eqnarray}
where we have used the canonical commutation relation
$[\hat{\vartheta}_0,\hat{\pi}_0] = i$. The operator $\hat{\Pi}_0$,
given by (\ref{label4.17}), coincides with the differential operator in
the l.h.s.\ of (\ref{label4.13}) in the $\vartheta_0$--representation.

The unitary transformation (\ref{label4.16}) is canonical, since it
retains the canonical commutation relations
\begin{eqnarray}\label{label4.18}
&&[\hat{\vartheta}_0, \hat{\Pi}_0] = \Big[U\,
\hat{\vartheta}_0\, U^{\dagger}, U\, \hat{\pi}_0\,
U^{\dagger}\Big] = [\hat{\vartheta}_0, \hat{\pi}_0] = i,\nonumber\\
&&[\hat{\Pi}_0, \hat{\Pi}_0] = \Big[U\, \hat{\pi}_0\,
U^{\dagger}, U\, \hat{\pi}_0\,
U^{\dagger}\Big] = [\hat{\pi}_0, \hat{\pi}_0] = 0.
\end{eqnarray}
According to Anderson \cite{AN94} the transformations (\ref{label4.15})
can be called  {\it similarity (gauge)} transformations.

Due to the canonical transformation (\ref{label4.14}) the field
operator $\hat{\vartheta}_0$ does not change but the Hamilton operator
transforms as follows
\begin{eqnarray}\label{label4.19}
\hat{h}_0(\hat{\vartheta}_0,\hat{\pi}_0) \to
\hat{H}_0(\hat{\vartheta}_0,\hat{\Pi}_0) =
U\,\hat{h}_0(\hat{\vartheta}_0,\hat{\pi}_0)\,
U^{\dagger} = \frac{\beta^2}{2L}\,\hat{\Pi}^2_0.
\end{eqnarray}
Equation (\ref{label4.13}) can be rewritten as 
\begin{eqnarray}\label{label4.20}
\hat{\Pi}_0\,{\langle \vartheta_0|\Omega(0)\rangle_{BCS}} = 0.
\end{eqnarray}
This means that the wave function (\ref{label4.7}) is the
eigenfunction of the momentum operator $\hat{\Pi}_0$ and the Hamilton
operator $\hat{H}_0$ with eigenvalue zero. 

The same result can be obtained using
\begin{eqnarray}\label{label4.21}
\hat{h}[\vartheta]|\Omega_0 \rangle &=&
\Big(\frac{\beta^2}{2L}\,\hat{\pi}^2_0 + \frac{2\pi}{L}\sum_{n \in
\mathbb{Z}}|n|\,a^{\dagger}_n a_n\Big)|\Omega_0 \rangle = 0,\nonumber\\
\hat{\pi}[\vartheta]|\Omega_0 \rangle &=&
\Big(\hat{\pi}_0 + \frac{2\pi}{L}\sum_{n \in
\mathbb{Z}}n\,a^{\dagger}_n a_n\Big)|\Omega_0 \rangle = 0.
\end{eqnarray}
By the canonical transformation (\ref{label4.16}) we transcribe
(\ref{label4.21}) as follows
\begin{eqnarray}\label{label4.22}
U\,\hat{h}[\vartheta]\,U^{\dagger} |\Omega(0)\rangle_{BCS}
&=&\Big(\frac{\beta^2}{2L}\,\hat{\Pi}^2_0 + \frac{2\pi}{L}\sum_{n \in
\mathbb{Z}}|n|\,a^{\dagger}_n a_n\Big)|\Omega(0)\rangle_{BCS} =
0,\nonumber\\ U\,\hat{\pi}[\vartheta]\,U^{\dagger}
|\Omega(0)\rangle_{BCS} &=& \Big(\hat{\Pi}_0 + \frac{2\pi}{L}\sum_{n
\in \mathbb{Z}}n\,a^{\dagger}_n a_n\Big)|\Omega(0)\rangle_{BCS} = 0.
\end{eqnarray}
This proves that the wave function $|\Omega(0)\rangle_{BCS}$
describes the ground state of the free massless (pseudo)scalar field
$\vartheta(x)$ defined by the Lagrangian (\ref{label1.1}). This is the
non--perturbative ground state describing the phase of the
spontaneously broken continuous symmetry (\ref{label1.2}) related to
the chiral symmetry of the massless Thirring model
\cite{FI1,FI2}. The wave functions (\ref{label4.10}) obey
the same equations (\ref{label4.22})
\begin{eqnarray}\label{label4.23}
&&\Big(\frac{\beta^2}{2L}\,\hat{\Pi}^2_0 + \frac{2\pi}{L}\sum_{n \in
\mathbb{Z}}|n|\,a^{\dagger}_n a_n\Big)|\Omega(\alpha)\rangle_{BCS} =
0,\nonumber\\ &&\Big(\hat{\Pi}_0 + \frac{2\pi}{L}\sum_{n \in
\mathbb{Z}}n\,a^{\dagger}_n a_n\Big)|\Omega(\alpha)\rangle_{BCS} = 0.
\end{eqnarray}
Finally, we would like to show that the ``magnetic'' quantum number $m$
defines the chirality of the fermionic state. In order to prove this
we suggest to use the results obtained by Nambu and Jona--Lasinio
\cite{YN60}. This concerns the analysis of the BCS wave function in
terms of the wave functions with a certain chirality $X$, the
eigenvalue of the $\gamma^5$ operator, $X = 0,\pm 1,\pm 2,\ldots $. The
BCS wave function of the ground state of the massless Thirring model
is defined by \cite{FI1,FI6}
\begin{eqnarray}\label{label4.24}
|\Omega(0)\rangle_{BCS} = \prod_{k^1}[u_{k^1} +
 v_{k^1}\,a^{\dagger}(k^1)b^{\dagger}(-k^1)]\,|\Psi_0\rangle,
\end{eqnarray}
where the coefficients $u_{k^1}$ and $v_{k^1}$ have the properties:
(i) $u^2_{k^1} + v^2_{k^1} = 1$ and (ii) $u_{- k^1} = u_{k^1}$ and
$v_{-k^1} = - v_{k^1}$ \cite{FI1,FI6}, $a^{\dagger}(k^1)$
and $b^{\dagger}(k^1)$ are creation operators of fermions and
antifermions with momentum $k^1$. According to Nambu and Jona--Lasinio
\cite{YN60} the wave function (\ref{label4.24}) should be a linear
superposition of the eigenfunctions $|\Omega_{2n}\rangle$ with
eigenvalues $X_n = 2n\,, n\in {\mathbb{Z}}$, i.e.
\begin{eqnarray}\label{label4.25}
|\Omega(0)\rangle_{BCS} =\sum_{n \in
 \mathbb{Z}}C_{2n}|\Omega_{2n}\rangle.
\end{eqnarray}
For chiral rotations of fermion fields with a chiral phase $\alpha_{\rm
A}$ the wave function (\ref{label4.24}) changes as follows
\cite{FI1}
\begin{eqnarray}\label{label4.26}
|\Omega(\alpha_{A})\rangle_{BCS} = \prod_{k^1}[u_{k^1} +
 v_{k^1}\,e^{\textstyle -2i\varepsilon(k^1)\alpha_{A}}\,
 a^{\dagger}(k^1)b^{\dagger}(-k^1)]\,|\Psi_0\rangle,
\end{eqnarray}
where $\varepsilon(k^1)$ is a sign function. In terms of
$|\Omega(\alpha_{A})\rangle$ the products $C_{2n} |\Omega_{2n}
\rangle$ are defined by
\begin{eqnarray}\label{label4.27}
C_{2n}|\Omega_{2n}\rangle =
 \int^{2\pi}_0\frac{d\alpha_{A}}{2\pi}\,e^{\textstyle +2in\alpha_{\rm
 A}}\,|\Omega(\alpha_{A})\rangle_{BCS}.
\end{eqnarray}
Substituting (\ref{label4.27}) in (\ref{label4.25}) and using the
identity \cite{IG64}
\begin{eqnarray}\label{label4.28}
\sum_{n\in \mathbb{Z}}e^{\textstyle\,2in\alpha_{\rm A}} =
\pi\sum_{k\in \mathbb{Z}}\delta(\alpha_{\rm A} - 2k\pi)
\end{eqnarray}
one arrives at the BCS wave function (\ref{label4.24}).
 
The bosonized version of the eigenfunctions $C_{2n}|\Omega_{2n}\rangle$
can be found in analogy with (\ref{label4.8}) and reads
\cite{FI6}
\begin{eqnarray}\label{label4.29}
&&C_{2n}\,|\Omega_{2n}\rangle \to
\int^{2\pi}_0\frac{d\alpha_{A}}{2\pi}\,e^{\textstyle +2in\alpha_{\rm
A}}\,e^{\textstyle \,i\lambda\sin(\hat{\vartheta}_0 - 2\alpha_{\rm
A})}|\Omega_0\rangle =\nonumber\\ &=& \sum_{m\in
\mathbb{Z}}J_m(\lambda)|m\rangle\otimes |\Psi_0\rangle
\int^{2\pi}_0\frac{d\alpha_{A}}{2\pi}\,e^{\textstyle +2i(n -
m)\alpha_{\rm A}} = J_n(\lambda)|n\rangle \otimes|\Psi_0\rangle.
\end{eqnarray}
This completes the proof. Hence, in the treatment of the collective
zero--mode as a rigid rotor the ``magnetic'' quantum number $m$
defines the chirality $X_m = 2m$ of the fermionic state in the
massless Thirring model.

The fact that the BCS wave function is not an eigenstate of chirality
testifies that chiral symmetry is spontaneously broken. In order to
clarify this assertion we would like to draw a similarity between
chirality in the massless Thirring model with triality in QCD. In QCD
there exist no triality changing transitions, this means a dynamical
change of triality is impossible. It is well--known that the confined
phase in QCD is $Z(3)$ symmetric. Triality zero states are screened,
and triality non--zero states are confined. This means that states
with different triality behave differently. Whereas in the
high--temperature phase of QCD all triality states behave in the same
way, they get screened. This is guaranteed by the spontaneous breaking
of $Z(3)$ symmetry. In our case the situation is similar to the
deconfined phase. In the massless Thirring model there are no
chirality changing transitions. The ground state is of BCS-type,
defining a condensate of fermion-antifermion pairs with different
chiralities. In order to get a ground state with properties
independent on the exact value of the total chirality of all
fermion--antifermion pairs, we need spontaneous breaking of chiral
symmetry similar to the the spontaneous breaking of $Z(3)$ symmetry in
QCD. Such a spontaneous breaking of chiral symmetry is realized by the
BCS wave function.

\section{Conclusion}

\hspace{0.2in} We have shown that the ground state of the free
massless (pseudo)scalar field, the bosonized version of the massless
Thirring model in the non--trivial phase, can be defined by a direct
product of the fiducial vacuum $|\Psi_0\rangle$ and a BCS--type wave
function (\ref{label4.1}). We have demonstrated that the BCS wave
function is related to the collective zero--mode described by a rigid
rotor (\ref{label4.8}). BCS wave functions differing in the values of
the field--shifts (\ref{label1.2}) are orthogonal ${_{BCS}\langle
\Omega(\alpha\,'\,)|\Omega(\alpha)\rangle_{BCS}}
=\delta_{\alpha\,\alpha}$.

We have analysed the generating functional of Green functions
$Z[J]$. We have shown that for $\tilde{J}(0)\neq 0$ the infrared
divergences have a simple physical interpretation in terms of a
classically moving rigid rotor acquiring an infinite angle for an
infinite interim even if its motion has been initiated by an
infinitesimal external perturbation. These divergences can be removed
by the constraint on the external source $\tilde{J}(0) = 0$.  As a
result the collective zero--mode cannot be excited and the correlation
functions are determined by the contribution of the vibrational modes
$\vartheta_{\rm v}(x)$ only. These modes are quantized relative to the
fiducial vacuum $|\Psi_0\rangle$.  According to \cite{FI3,FI4} this
testifies that the quantum field theory of the free massless
(pseudo)scalar field $\vartheta_{\mathrm v}(x)$, in the Wightman sense
\cite{AW64}, deals with Wightman's observables defined on the test
functions from ${\cal S}_0(\mathbb{R}^{\,2})$.

The BCS type wave function (\ref{label4.8}) of the ground state is not
invariant under the continuous symmetry (\ref{label1.2}) and behaves
as (\ref{label4.10}). Hence, according to the Goldstone theorem
\cite{JG61} the continuous symmetry (\ref{label1.2}) is spontaneously
broken. As has been shown in \cite{FI2,FI3,FI4} the
phase of spontaneously broken continuous symmetry (\ref{label1.2}) is
characterized quantitatively by the non--vanishing spontaneous
magnetization ${\cal M} = \langle \Psi_0| \cos\beta\vartheta_{\mathrm
v}(x) |\Psi_0 \rangle = 1$. This confirms the non--vanishing value of
the fermion condensate in the massless Thirring model with fermion
fields quantized in the chirally broken phase \cite{FI1}. Hence,
the massless Thirring model possesses a chirally broken phase as has
been pointed out in
\cite{FI1}--\cite{FI4,FI6,FI7}.

In the rigid rotor treatment of the collective zero--mode the
variation $\delta \vartheta(x)$ of the free massless (pseudo)scalar
field $\vartheta(x)$, caused by the field--shift transformation
(\ref{label1.2}), is defined by a canonical quantum mechanical
commutator
$$
\delta \vartheta(x) = \alpha\,i[Q(x^0),\vartheta(x)] =
\alpha\,i[\pi_0,\vartheta_0] = \alpha,
$$
which can never be equal to zero \cite{SC73}. This result does not
depend on whether the ground state of the free massless (pseudo)scalar
field is invariant or non--invariant under symmetry transformations
(\ref{label1.2}).

Since the removal of the collective zero--mode from the observable
modes of the free massless (pseudo)scalar field $\vartheta(x)$ agrees
with the definition of Wightman's observable on the test functions
from ${\cal S}_0(\mathbb{R}^{\,2})$ the obtained non--vanishing of the
variation $\delta \vartheta(x)$ does not contradict Coleman's theorem
valid only for Wightman's observables defined on the test functions
from ${\cal S}(\mathbb{R}^{\,2})$ \cite{FI3,FI4}.


\newpage

\begin{thebibliography}{9}
\bibitem{FI1} 
M. Faber and A. N. Ivanov, 
Eur. Phys. J. C {\bf 20}, 723
(2001), hep-th/0105057.
\bibitem{FI2} 
M. Faber and A. N. Ivanov, 
Eur. Phys. J. C {\bf 24}, 653 (2002), hep--th/0112184.
\bibitem{FI3}
M. Faber and A. N. Ivanov,
{\it On spontaneous breaking of continuous symmetry in
1+1--dimensional space--time}, hep--th/0204237.
\bibitem{FI4}
M. Faber and A. N. Ivanov,
{\it Quantum field theory of a free massless (pseudo)scalar field in
1+1--dimensional space--time as a test for the massless Thirring model},
hep--th/0206244.
\bibitem{KY76}
K. Yoshida,
Nucl. Phys. B {\bf 105}, 272 (1976).
\bibitem{FI5}
M. Faber and A. N. Ivanov,
{\it On the solution of the massless
Thirring model  with fermion
fields quantized in the chiral symmetric phase},
hep--th/0112183.
\bibitem{MWH} 
N. D. Mermin and H. Wagner, 
Phys. Rev. Lett. {\bf 17},
1133 (1966); 
P. C. Hohenberg, 
Phys. Rev. {\bf 158}, 383 (1967);
N. D. Mermin, J. Math. Phys. {\bf 8}, 1061 (1967).
\bibitem{SC73} 
S. Coleman, 
Comm. Math. Phys. {\bf 31}, 259 (1973).
\bibitem{AW64} 
A. S. Wightman, 
{\it Introduction to Some Aspects of
the Relativistic Dynamics of Quantized Fields}, in 
{\it HIGH ENERGY ELECTROMAGNETIC INTERACTIONS AND FIELD THEORY}, 
Carg$\grave{\rm e}$se
Lectures in Theoretical Physics, edited by M. Levy , 1964, Gordon and
Breach, 1967, pp.171--291;
R. F. Streater and A. S. Wightman,
in {\it PCT, SPIN AND STATISTICS, AND ALL THAT},
Princeton University Press, Princeton and Oxford, Third Edition, 1980.
\bibitem{JG61}
J. Goldstone,
Nuovo Cimento {\bf 19}, 154 (1961);
J. Goldstone, A. Salam, and S. Weinberg,
Phys. Rev. {\bf 127}, 965 (1962).
\bibitem{SC75}
S. Coleman,
Phys. Rev. D {\bf 11}, 2088 (1975).
\bibitem{IG64}
I. M. Gel'fand and G. E. Shilov,
in {\it GENERALIZED FUNCTIONS, Properties and Operations}, 
Vol.1, Academic Press, New York, 1964.
\bibitem{JW80} 
A. L. Fetter and J. D. Walecka,
in {\it THEORETICAL MECHANICS OF PARTICLES AND CONTINUA},
McGraw--Hill Publishing Co., New York, 1980, pp.86--130;
K. Huang,
in {\it QUANTUM FIELD THEORY, From Operators to Path Integrals},
John Willey $\&$ Sons, Inc., New York, 1998, pp.1--7.
\bibitem{KS75}
J. Kogut and L. Susskind,
Phys. Rev. D {\bf 11}, 3594 (1975).
\bibitem{SW70}
J. Schwinger 
in {\it PARTICLES AND SOURCES}, Gordon and
Breach, New York 1969 and 
{\it PARTICLES, SOURCES AND FIELDS},
Addison--Wesley Publishing Co., Massachusetts 1970
\bibitem{SH81}
L. S. Schulman,Phys. Rev. {\bf 176}, 1558 (1968); 
Phys. Rev. {\bf 188}, 1139 (1969);
J. Math. Phys. {\bf 12}, 304 (1971);
L. S. Schulman,
in {\it TECHNIQUES AND APPLICATIONS OF PATH INTEGRATION},
John $\,\&\,$ Sons, New York, 1981, pp.190--213.
\bibitem{FI6}
M. Faber and A. N. Ivanov,
Phys. Lett. B {\bf 563}, 231 (2003).
\bibitem{AS72}
{\it HANDBOOK OF MATHEMATICAL FUNCTIONS, with Formulas,
Graphs, and Mathematical Tables}, ed. by M. Abramowitz and
I. E. Stegun, U.S. Department of Commerce, National Bureau of
Standards, Applied Mathematics Series $\bullet$ 55, 1972,
p.361 formula (9.1.41).
\bibitem{GR65}
I. S. Gradshtein and I. M. Ryzhik,
in {\it TABLE OF INTEGRALS, SERIES AND PRODUCTS},
Academic Press, New York and London, 1965, p.980 formula (8.536.3).
\bibitem{AS72a}
(see \cite{AS72} p.979 formula (8.531.3)).
\bibitem{AN94}
A. Anderson,
Ann. of Phys. {\bf 232}, 292 (1994).
\bibitem{YN60}
Y. Nambu and G. Jona--Lasinio,
Phys. Rev. {\bf 122}, 345 (1960).
\bibitem{FI7}
M. Faber and A. N. Ivanov, 
J. of Phys. A {\bf 36}, issue 28 (2003), hep--th/0205249;
{\it Massless Thirring model in the boson field representation}.
hep--th/0206034.
\end{thebibliography}
\end{document}